# Comment: Fisher Lecture: Dimension Reduction in Regression

**Lexin Li and Christopher J. Nachtsheim**

## 1. INTRODUCTION

Professor Cook is to be congratulated for his ground-breaking work in dimension reduction in regression. The paper develops a general theoretical foundation for studying principal components and other dimension reduction methods in a regression context. This framework yields a basis for elucidating the strengths, weaknesses and relationships among the various dimension reduction methods, including ordinary least squares (OLS), principal components regression (PCR), sliced inverse regression (SIR), parametric inverse regression and partial least squares. The promising new method, principal fitted components (PFC), appears to outperform some long-standing approaches such as PCR, OLS and SIR. Finally, as a result of this contribution, the standard approach to regression, with its emphasis on fixed predictors and the need to assume away the randomness of $\mathbf{X}$, and the standard approach to principal components, with its focus on the correlation matrix rather than the covariance matrix, both seem to be under question.

Specific contributions of Professor Cook's paper include the following: (1) It provides a theoretical foundation for the widely used principal components regression. (2) It resorts to a model and thus a likelihood function, through the inverse regression of predictors given response, to study sufficient reduction in a forward regression problem. Consequently, likelihood-based inferences can be developed, and the inferential capabilities of dimension reduction are moved closer to mainstream regression methodology. (3) It permits extension to categorical or mixtures of continuous and categorical predictors, an area that most existing model-free dimension reduction approaches do not handle effectively. (4) It seems to be applicable to problems where the number of predictors exceeds the number of observational units, which is perhaps one of the most challenging current frontiers in statistical methodology. In general, we believe that this paper has paved the way for substantive research in dimension reduction, and that it will surely be the subject of much future application and elaboration.

In what follows, we focus on two issues raised but not thoroughly addressed in Professor Cook's paper: (1) the role of predictor screening and a connection with the recently proposed supervised principal components method and (2) dimension reduction in the presence of binary predictors.

## 2. SUPERVISED PRINCIPAL COMPONENTS

Professor Cook has suggested a combination of predictor screening and principal fitted components analysis when the number of predictors $p$ is large—in particular when $n < p$. As Professor Cook noted, traditional studies have based predictor screening on the univariate forward regressions of $Y$ on individual $X_j$, $j = 1, \ldots, p$. In view of the results presented in this paper, it appears that the screening at the outset should probably be based on the univariate inverse regressions of $X_j$ on $\mathbf{f}_y$. This in turn suggests the following algorithm for PFC in conjunction with predictor screening:

1. Compute the inverse univariate regressions $X_j$ on $\mathbf{f}_y$, for $j = 1, \ldots, p$.
2. Form a reduced matrix $\mathbf{X}_\theta$ composed of only those predictors whose regression coefficient on $\mathbf{f}_y$ is determined to surpass a level-of-significance threshold $\theta$.
3. Obtain $\widehat{\boldsymbol{\Gamma}}$ using PFC based on the reduced $\mathbf{X}_\theta$.

*Lexin Li is Assistant Professor, Department of Statistics, North Carolina State University, Box 8203, Raleigh, North Carolina 27695, USA e-mail: li@stat.ncsu.edu. Christopher J. Nachtsheim is Professor, School of Statistics, University of Minnesota, 224 Church Street S. E., Minneapolis, Minnesota 55455, USA e-mail: cnachtsheim@csom.umn.edu.*

4. Pass $\widehat{\boldsymbol{\Gamma}}$ to the forward regression.

Recently, Bair et al. (2006) proposed an approach called supervised principal components (SPC), which can be described as follows:

1. Compute the forward univariate regressions $Y$ on individual $X_j$, $j = 1, \ldots, p$.
2. Form a reduced $\mathbf{X}_\theta$ matrix composed of only those features whose regression coefficient exceeds a threshold $\theta$ in absolute value.
3. Compute the first, or first few, principal components of the reduced data matrix.
4. Use these principal component(s) in a regression model to predict the outcome.

The two approaches look strikingly similar, but there are some essential differences. First, with Cook's approach, predictor screening is based on univariate inverse regressions, whereas forward univariate regressions are employed in SPC. Of course, the two screening approaches would frequently yield similar reductions since the significance of a regression of $Y$ on $X_j$ will often be associated with significance in the regression of $X_j$ on $Y$. Differences between two regressions arise in some cases, however. For instance, if $X_j | Y$ is quadratic in $Y$ with no linear trend, $X_j$ would be retained by an inverse quadratic regression but deleted by the forward SPC screening, since there would be no linear trend in $Y|X_j$. Conversely, if $Y|X_j$ is quadratic in $X_j$, again with no linear trend, it will be deleted by both procedures. This is because the forward screening considers only linear trend (and there is none) while the inverse regression $X_j|Y$ would exhibit a null trend with nonconstant variance.

Second, we note that the underlying model assumed by SPC can be viewed as a special case of the PFC model given in Cook (2007). Consider (9) and (10) of Bair et al. (2006), in which both the response and the predictors are linked by a univariate latent variable $U$:

$$Y = \beta_0 + \beta_1 U + \varepsilon,$$
$$X_j = \alpha_{0j} + \alpha_{1j} U + \epsilon_j, \quad j = 1, \ldots, p,$$

where $\varepsilon$ and $\epsilon_j$ are assumed to have mean 0, are independent and are also independent of $U$. Consequently,

$$X_j | (Y = y)$$
$$= \left(\alpha_{0j} - \frac{\beta_0 \alpha_{1j}}{\beta_1}\right) + \frac{\alpha_{1j}}{\beta_1} y + \left(\epsilon_j - \frac{\alpha_{1j}}{\beta_1} \varepsilon\right)$$
$$\equiv \mu_j + \gamma_j y + \varepsilon_j^*.$$

Let $\boldsymbol{\Gamma}$ denote the $p \times 1$ vector with the $j$th element $\gamma_j/c$, where $c = (\sum_{j=1}^p \gamma_j^2)^{1/2}$, $\mathbf{f}_y = c(y - \bar{y})$, $\boldsymbol{\mu}$ denotes the $p \times 1$ vector with the $j$th element $\mu_j + \gamma_j \bar{y}$ and $\boldsymbol{\varepsilon}^*$ denotes the $p \times 1$ vector with the $j$th element $\varepsilon_j^*$. We then obtain

(1) $$\mathbf{X}_y = \boldsymbol{\mu} + \boldsymbol{\Gamma} \mathbf{f}_y + \boldsymbol{\varepsilon}^*.$$

Model (1) is similar to model (5) in Cook (2007), where the predictors $X_j$ given $Y$ are conditionally independent since $\varepsilon$ and $\epsilon_j$ are independent, although $\operatorname{var}(X_j | Y = y)$ are not necessarily equal. Let $\sigma_\varepsilon^2 = \operatorname{var}(\varepsilon)$, $\boldsymbol{\epsilon}_x = (\epsilon_1, \ldots, \epsilon_p)^\mathsf{T}$, and $\Sigma_{\boldsymbol{\epsilon}_x} = \operatorname{var}(\boldsymbol{\epsilon}_x)$, the diagonal matrix with the $j$th diagonal element equal to $\operatorname{var}(\epsilon_j)$. Then some algebra reveals that the model (1) can be viewed in the form of model (13) in Cook (2007),

$$\mathbf{X}_y = \boldsymbol{\mu} + \boldsymbol{\Gamma} \mathbf{f}_y + \boldsymbol{\Gamma}_0 \boldsymbol{\Omega}_0 \boldsymbol{\varepsilon}_0 + \boldsymbol{\Gamma} \boldsymbol{\Omega} \boldsymbol{\varepsilon},$$

where $\boldsymbol{\Gamma}_0 \in \mathbb{R}^{p \times (p-1)}$ denotes an orthogonal completion of $\boldsymbol{\Gamma}$, $\boldsymbol{\Omega}_0 = (\boldsymbol{\Gamma}_0^\mathsf{T} \Sigma_{\boldsymbol{\epsilon}_x} \boldsymbol{\Gamma}_0)^{1/2}$, $\boldsymbol{\Omega} = (\boldsymbol{\Gamma}^\mathsf{T} \Sigma_{\boldsymbol{\epsilon}_x} \boldsymbol{\Gamma} + c^2 \sigma_\varepsilon^2)^{1/2}$, $\boldsymbol{\varepsilon}$ is an error variable with mean 0 and variance 1, $\boldsymbol{\varepsilon}_0$ is a $(p-1) \times 1$ vector of error variables with mean $\mathbf{0}$ and identity variance, $\boldsymbol{\varepsilon}$ and $\boldsymbol{\varepsilon}_0$ are independent, and $\boldsymbol{\mu}$, $\boldsymbol{\Gamma}$ and $\mathbf{f}_y$ are as defined before. Clearly, the models discussed in Cook (2007) are more flexible than model (1) assumed by SPC, in the sense that they permit a more flexible choice for $\mathbf{f}_y$ and a more flexible variance structure.

To compare the two approaches empirically, we applied SPC and PFC to the logo data of Henderson and Cote (1998), which was also examined by Li and Nachtsheim (2006) in the context of dimension reduction in regression. The data characterize $n = 195$ company logos in terms of their overall affect, $Y$, which is a composite of various subjective reactions to a logo and 22 design characteristics $X$. Using a significance level of $\alpha = 0.05$ for predictor screening, univariate forward regressions led to the dropping of eight predictors for the SPC approach, while the univariate inverse regressions on $\mathbf{f}_y = (y - \bar{y}, y^2 - \bar{y}^2)^\mathsf{T}$ implicated the same eight predictors plus one other. Since it was concluded in Li and Nachtsheim (2006) that the data exhibit one-dimensional structure $(d = 1)$, we employed SPC and PFC to extract the first reduction variable $\widehat{\boldsymbol{\Gamma}}^\mathsf{T} \mathbf{X}$. Figure 1 shows the summary plot of $Y$ versus $\widehat{\boldsymbol{\Gamma}}^\mathsf{T} \mathbf{X}$ with and without predictor screening. For the full predictor case, PFC chose the fifth PC direction, whereas SPC is always based on the first. Nonetheless, the results are quite similar. Following screening, PFC selected the first PC direction. The similarities in the plots in this case are as expected.

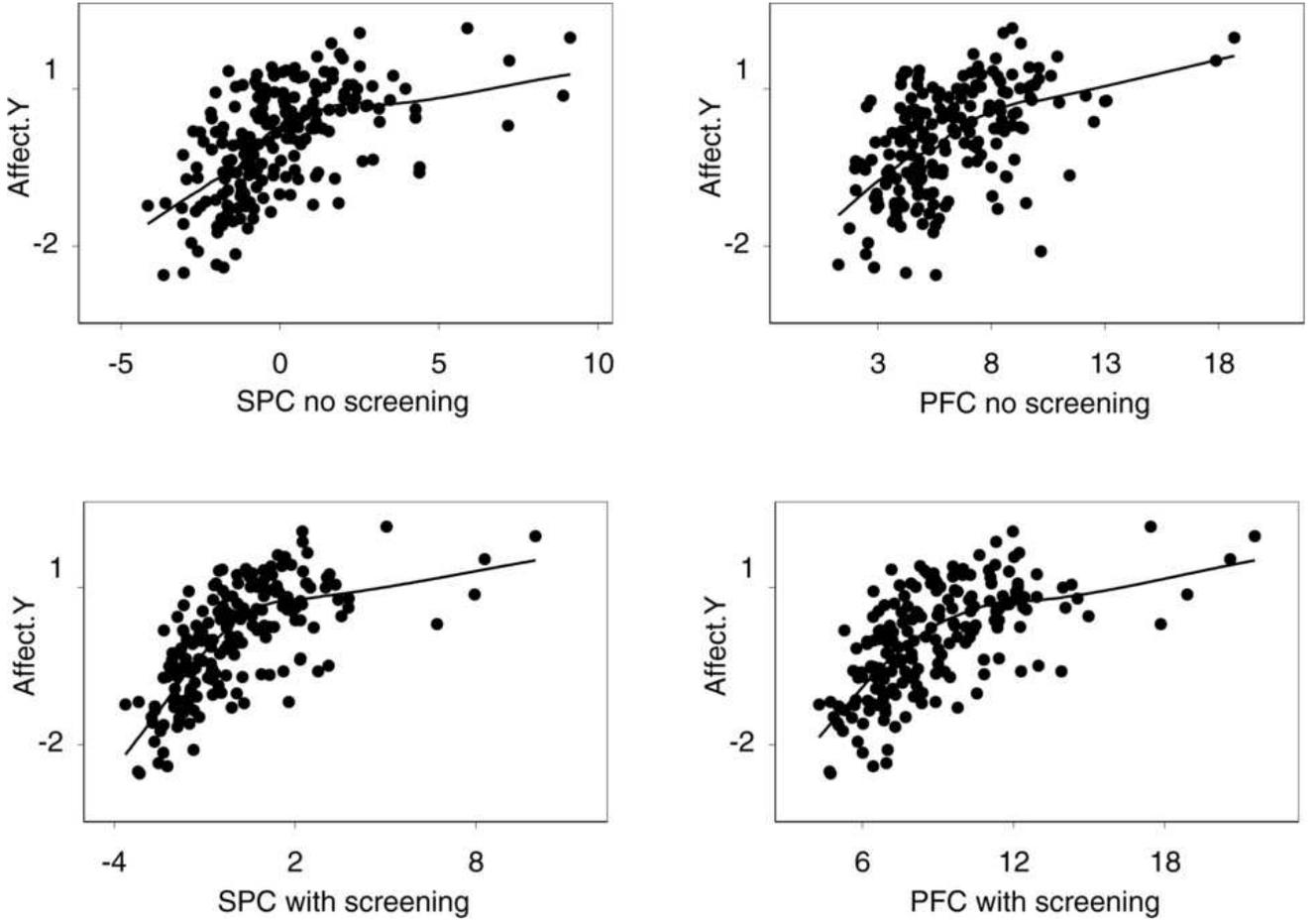

Fig. 1. *Sufficient summary plots for SPC and PFC with and without predictor screening.*

## 3. BINARY PREDICTORS

A key contribution of Professor Cook's paper has been to provide the logic on which to base an extension of dimension reduction to categorical predictors. Section 3.3 of Cook (2007) gives a brief prescription for such extensions, and we focus here on the use of principal components for binary predictors.

In many applications such as in genetics and data mining, it is common to encounter data sets where all predictors are dichotomous. For instance, in quantitative genetics, the genetic architecture is recorded by a sequence of binary markers. In a typical email spam filtering system, each email is denoted by a vector of binary features indicating whether a set of commonly occurring words is present. These examples underscore the relevance of binary predictors in applications, yet existing dimension reduction approaches have focused mainly on the continuous-predictor case.

Following the formulation given in Section 3.3 of Cook (2007), we assume that each predictor $X_j$, $j = 1, \ldots, p$, given $Y = y$, follows a Bernoulli distribution with parameter $p_j(y) = \Pr(X_j = x_j|y)$, and all predictors are conditionally independent. The log likelihood for a single $x_j$ at $y$, $x_{jy}$, is therefore

$$(2) \qquad \log q_j(y) + x_{jy}(\mu_j + \gamma_j^\mathsf{T} \boldsymbol{\nu}_y),$$

where $q_j(y) = 1 - p_j(y) = 1/\{1 + \exp(\mu_j + \gamma_j^\mathsf{T} \boldsymbol{\nu}_y)\}$. The overall log likelihood is then

$$(3) \qquad \sum_y \sum_j \{\log q_j(y) + x_{jy}(\mu_j + \gamma_j^\mathsf{T} \boldsymbol{\nu}_y)\}.$$

To obtain the maximum likelihood estimate of $\boldsymbol{\mu}$, $\boldsymbol{\nu}$ and $\boldsymbol{\Gamma}$, an iterative algorithm can be employed. More specifically, fixing $\boldsymbol{\mu}$ and $\boldsymbol{\Gamma}$, we first estimate the $n$ $\boldsymbol{\nu}_y$'s by adding the log likelihood (2) over $j$ and obtaining $\sum_j \{\log q_j(y) + x_{jy}(\mu_j + \gamma_j^\mathsf{T} \boldsymbol{\nu}_y)\}$. Given $\mu_j$ and $\gamma_j$, estimates can be obtained from a standard logistic regression maximization, by treating $(x_{1y},$ $\dots, x_{py})$ as the "response," $(\mu_1, \dots, \mu_p)$ as the "offset" and $\gamma_j^\mathsf{T}$ as the $j$th value of a $d$-dimensional vector of "predictors," and the logistic regression is fitted through the origin. We estimate the $d$-dimensional parameter vector $\boldsymbol{\nu}_y$ based on $p$ observations, and in total we fit $n$ logistic regressions. Next, to estimate $\mu_j$ given fixed $\boldsymbol{\nu}_y$ and $\boldsymbol{\Gamma}$, we simply maximize $\sum_y \{\log q_j(y) + x_{jy}(\mu_j + \gamma_j^\mathsf{T}\boldsymbol{\nu}_y)\}$. This corresponds to a univariate logistic regression with intercept $\mu_j$ and offset $\gamma_j^\mathsf{T}\boldsymbol{\nu}_y$, and we fit $p$ such logistic regressions in total. Finally, fixing $\boldsymbol{\mu}$ and $\boldsymbol{\nu}$, we need to maximize (3) over $\boldsymbol{\Gamma}$ in a Grassmann manifold. Optimization in Grassmann manifolds is discussed in Edelman, Arias and Smith (1999).

We implemented the above iterative optimization algorithm and tested the method with the following simulation setup. The response $Y$ was generated as a normal random variable with mean 0 and variance $\sigma_Y^2$, and $\mathbf{X}|Y = y$ was generated as a Bernoulli random vector with the natural parameter $\boldsymbol{\eta}_y \in \mathbb{R}^p$ and the canonical link function $g$,

$$\boldsymbol{\eta}_y = g(\mathrm{E}(\mathbf{X}|Y=y)) = \boldsymbol{\Gamma}\boldsymbol{\beta}\mathbf{f}_y.$$

We fixed the sample size $n = 200$ and the number of predictors $p = 20$. We first considered the $d = 1$ case and examined different forms of $\boldsymbol{\Gamma}$: $\boldsymbol{\Gamma}_1 = (1, \dots, 1, 0, \dots, 0)^\mathsf{T}/\sqrt{10}$, with ten 1s and the remaining 0s; $\boldsymbol{\Gamma}_2 = (1, 1, 1, 0.5, 0.5, -0.5, -0.5, -1, -1, -1, 0, \dots, 0)^\mathsf{T}/\sqrt{7}$; and $\boldsymbol{\Gamma}_3 = (1, \dots, 1, 0.5, \dots, 0.5, -0.5, \dots, -0.5, -1, \dots, -1)^\mathsf{T}/\sqrt{12.5}$, with each element repeated five times. $\boldsymbol{\beta} = 1$ and $\mathbf{f}_y$ is the centered linear term $y$. We also considered a $d = 2$ case, by choosing $\boldsymbol{\Gamma}_4 = ((1, \dots, 1, 0, \dots, 0)^\mathsf{T}, (0, \dots, 0, 1, \dots, 1)^\mathsf{T})/\sqrt{10}$, with ten 1s in each direction; $\boldsymbol{\beta} = \mathrm{diag}(1, 0.1)$; and $\mathbf{f}_y = (y, y^2)^\mathsf{T}$ centered.

Table 1 reports the average angles (in degrees) between $\boldsymbol{\Gamma}$ and $\widehat{\boldsymbol{\Gamma}}$ out of 50 data replications, as $\sigma_Y$, the term that controls the "signal" strength, increases. It is first noted that the estimation accuracy increases as $\sigma_Y$ increases, as expected. Second, we have observed that the optimization algorithm may become quite unstable in some cases, as reflected by the nonconvergence of the logistic fit when estimating $\boldsymbol{\mu}$ and $\boldsymbol{\nu}$. [This instability in turn has caused poor estimation accuracy. For instance, in the case of $\boldsymbol{\Gamma}_4$, even with a relatively large $\sigma_Y$ value, the average angle is large.]

To make the principal components for all binary predictors practically useful, a stable version of the iterative maximization algorithm is needed. One possible solution is the majorization strategy as suggested by Kiers (2002). Future research along this direction seems warranted.

TABLE 1
*Average of angles (in degrees) between $\boldsymbol{\Gamma}$ and $\widehat{\boldsymbol{\Gamma}}$ given $\sigma_Y$ based on 50 data replications*

| $\sigma_Y$ | 0.1 | 1 | 3 | 5 | 10 |
|---|---|---|---|---|---|
| $\boldsymbol{\Gamma}_1$ | 76.163 | 62.756 | 24.141 | 12.127 | 7.366 |
| $\boldsymbol{\Gamma}_2$ | 81.175 | 76.659 | 40.073 | 18.380 | 11.313 |
| $\boldsymbol{\Gamma}_3$ | 79.566 | 76.501 | 49.320 | 23.501 | 20.156 |
| $\boldsymbol{\Gamma}_4$ | 85.879 | 85.672 | 79.554 | 65.669 | 46.326 |